\documentclass[twocolumn,aip,reprint]{revtex4-2}

\usepackage{bm}
\usepackage{graphicx}
\usepackage{xcolor}
\usepackage{booktabs}
\usepackage{array}
\usepackage[hidelinks]{hyperref}

\usepackage{amsmath}
\usepackage{amssymb}

\newcommand{\ket}[1]{\lvert #1\rangle}
\newcommand{\bra}[1]{\langle #1\rvert}

\newcommand{\anticomm}[2]{\{#1,#2\}}
\newcommand{\Eq}[1]{Eq.~(\ref{#1})}

\newcommand{\Sec}[1]{Sec.~\ref{#1}}

\newcommand{\bsube}{\begin{subequations}}
\newcommand{\esube}{\end{subequations}}

\newcommand{\be}{\begin{equation}}
\newcommand{\ee}{\end{equation}}
\newcommand{\dg}{\dagger}
\newcommand{\la}{\langle}
\newcommand{\ra}{\rangle}

\begin{document}

\title{Correlated memory effect of environment in radical-pair magnetoreception
}

\author{Shixuan Chen}
\affiliation{State Key Laboratory of Precision and Intelligent Chemistry  and Department of Chemical Physics, University of Science and Technology of China, Hefei, Anhui 230026, China}

\author{Yao Wang}
 \email{wy2010@ustc.edu.cn}
\affiliation{Hefei National Research Center for Physical Sciences at the Microscale and Department of Chemical Physics, University of Science and Technology of China, Hefei, Anhui 230026, China}

\author{Rui-Xue Xu}
\email{rxxu@ustc.edu.cn}

\affiliation{Hefei National Research Center for Physical Sciences at the Microscale and Department of Chemical Physics, University of Science and Technology of China, Hefei, Anhui 230026, China}

\author{YiJing Yan}
\affiliation{Hefei National Research Center for Physical Sciences at the Microscale and Department of Chemical Physics, University of Science and Technology of China, Hefei, Anhui 230026, China}

\date{\today}

\begin{abstract}
In the
photoreceptor compass model, a light-generated, spin-correlated radical pair
undergoes singlet--triplet interconversion driven by Zeeman and anisotropic
hyperfine interactions, leading to 
orientation-dependent reaction yields. In this study, we consider simultaneously the spin environment and the boson environment surrounding the radical-pair system and concentrate on the correlated memory effect of the spin bath. 
In the linear-response limit, the spin bath is
mapped to an effective Gaussian environment. The system dynamics and yields under its influence together with the bosonic bath are propagated uniformly using the hierarchical equations of motion (HEOM)
or equivalently the dissipaton equations of
motion (DEOM) method. 
The HEOM/DEOM is non-Markovian and non-perturbative, exact for Gaussian environments. Corresponding results of the Markovian Lindblad master equation are also shown for comparison. Numerical demonstrations  highlight the non-Markovian memory effect as a crucial ingredient for radical-pair
magnetoreception. 

\end{abstract}

\maketitle

\section{Introduction}
\label{sec:intro}

Many animals, especially migratory birds, use the geomagnetic field as a source of directional information.\cite{Hab77165,Sch781,Tim9871,Rit00707,Joh05705,Rod09353,Hor16299,Fay18064107,Mou1850,Wil1920190295,Xu21535,Zad2220220325,Bra23111,Kry25893,Lis26985}
Among the proposed mechanisms, the radical-pair magnetoreception is a leading explanation for the light-dependent magnetic compass.
The basic picture is as follows.
Photoexcitation or photoinduced electron transfer generates a radical pair with correlated unpaired-electron spins, typically in an electronic singlet state.
During its finite lifetime, the radical pair evolves under the combined action of the geomagnetic field and local hyperfine interactions, which drive coherent interconversion between the singlet and triplet states. 
If these spin states undergo different reaction pathways, the spin evolution changes the relative yields of the corresponding chemical products.
This mechanism differs fundamentally from a classical compass. It does not require a permanent molecular magnetic moment to overcome thermal fluctuations and align with the geomagnetic field.
Instead, the weak geomagnetic field perturbs the spin dynamics before recombination and thereby modulates the chemical reaction outcome.

Cryptochrome is a natural molecular candidate for magnetoreception because it
is a blue-light-sensitive flavoprotein that can undergo photoinduced electron
transfer.\cite{Rit00707,Bis09404}
Its light-absorbing component is the flavin adenine dinucleotide (FAD)
cofactor.
Excitation of FAD is
followed by electron transfer along a chain
of nearby tryptophan residues, generating an initially
spin-correlated flavin--tryptophan radical pair, typically consisting
of a flavin radical and a tryptophan radical.\cite{Bis09404,Sol143845,Mul1515502}
Because the subsequent spin evolution of the radical pair can be influenced
by an external magnetic field, the yields of the reaction products may depend
on the field strength and direction.\cite{Rit00707,Mae08387}
In vitro experiments on cryptochrome 4 from the European robin (ErCRY4)
demonstrated that its photochemical reactions are magnetically sensitive and
clarified how successive tryptophan residues contribute to the
electron-transfer pathway.\cite{Xu21535}
These findings provide molecular support for cryptochrome-mediated
radical-pair magnetoreception.

In Ref.\,\onlinecite{Rit00707}, a radical-pair model is used to describe the
magnetic-field effect in avian magnetoreception.
In this model, the radical pair is initially in the singlet state,
and the singlet--triplet conversion is driven by the Zeeman interaction and
the hyperfine interactions between the electron and nuclear spins.
For simplicity, each radical contains one electron spin and one nuclear spin,
and an anisotropic hyperfine coupling is introduced to generate the
orientation dependence on the external magnetic field.
The singlet and triplet states then recombine through different reaction
channels, and the corresponding reaction yields are used to characterize the
orientational sensitivity of the radical pair.

Some previous studies have investigated the effects of environmental fluctuation,
spin relaxation, and decoherence on radical-pair dynamics using
open-system approachesl including Redfield theory, Born--Markov master
equations, stochastic Schr\"odinger equations, and phenomenological relaxation
models.\cite{Car15012720,Wor16035104,Ada1815719,Fay21084121,Gol23105102,Luo24074306}
In this study, we concentrate on a non-Markovian environment of spins coupled to this
radical-pair model.
The spin environment induces relaxation and dephasing of the radical pair,
while its finite memory and correlation between coupling modes may also influence the singlet--triplet conversion
and the subsequent reaction yields.
In a protein environment, however, those nuclear-spin
influences coexist with collective vibrational and conformational modes.
We therefore actually need to consider two non-Markovian environments acting simultaneously on
the radical pair: a spin bath and a bosonic bath.  The spin bath is
coupled through local  electron-spin operators while the effect of bosonic bath is taken to account for singlet--triplet energy 
fluctuation.
We employ a unified hierarchical equations of motion (HEOM)
\cite{Tan906676,Yan04216,Xu05041103,Shi09084105,Shi09164518,Tan20020901} or equivalently dissipaton equations of
motion (DEOM) framework to propagate the simultaneous influence of the two
baths in a nonperturbative and non-Markovian manner.
\cite{Yan14054105,Xu151816,Wan22170901,Yin24144112}
Within this framework, we investigate the effects of non-Markovian environmental fluctuations on radical-pair dynamics,
spin-selective reaction yields, and magnetic-field sensitivity.

The remainder of this paper is organized as follows.
In Sec.~\ref{thsec2}, we briefly introduce the biological background of avian
magnetoreception and present the radical-pair model considered in this work.
In Sec.~\ref{thsec3}, we formulate the mixed HEOM/DEOM approach for
simultaneous non-Markovian spin and bosonic baths and introduce a Markovian
reference,  the Lindblad master equation for comparison, to highlight the non-Markovian memory effect.
In Sec.~\ref{thsec4}, we present the numerical results and discuss the effects of the two environments on radical-pair dynamics,
reaction yields, and magnetic sensitivity.
Finally, Sec.~\ref{thsum} summarizes the main conclusions of this work.

\section{Open radical-pair model}
\label{thsec2}

We begin by considering the isolated radical-pair model introduced in the
Introduction. A minimal reaction scheme can be written as
\begin{align}
&[\mathrm{D}\mathrm{A}]^{\ast}
\xrightarrow{\text{ET}}
  [\bullet \mathrm{D}^{ +}\bullet \mathrm{A}^{ -}]^{S}
  \rightleftarrows
  [\bullet \mathrm{D}^{ +}\bullet \mathrm{A}^{-}]^{T},
  \nonumber\\
  &[\bullet \mathrm{D}^{ +}\bullet \mathrm{A}^{-}]^{S}
  \xrightarrow{k_S} \text{singlet products},
  \nonumber\\
  &[\bullet \mathrm{D}^{ +}  \bullet \mathrm{A}^{-}]^{T}
  \xrightarrow{k_T} \text{triplet products}.  \nonumber  
\end{align}
Here, $\mathrm{D}$ and $\mathrm{A}$ denote an electron donor and acceptor,
respectively, and ET denotes photoinduced electron transfer. The ET step generates an initially spin-correlated radical pair. Subsequent hyperfine and Zeeman interactions drive
singlet--triplet interconversion before the radical pair undergoes
spin-selective recombination with rate constants $k_S$ and $k_T$.

\subsection{Hamiltonian and recombination yields}

All electron and nuclear spin operators are taken to be dimensionless, with
spin-$1/2$ eigenvalues $\pm 1/2$ and
\begin{align}
 [\hat S_{j\alpha},\hat S_{j\beta}]
 =i\epsilon_{\alpha\beta\gamma}\hat S_{j\gamma},
 \qquad
 [\hat I_{j\alpha},\hat I_{j\beta}]
 =i\epsilon_{\alpha\beta\gamma}\hat I_{j\gamma}.
\nonumber
\end{align}
The nuclear
spins are modeled as two hydrogen nuclei.
We define
$\gamma_e=g\mu_B/\hbar$,
where $g\simeq2$ and $\mu_B$ is the Bohr magneton. The radical-pair Hamiltonian is
\begin{align}
 \hat H_{\rm RP}(\bm B)
 =\hbar\gamma_e\sum_{j=1}^{2}
 \hat{\bm S}_j\cdot
 \left(\bm B+\bm A_j\cdot\hat{\bm I}_j\right).
 \label{eq:HRP}
\end{align}
The first term in \Eq{eq:HRP} accounts for the electronic Zeeman interaction, while the second term accounts for electron--nuclear hyperfine coupling. $\bm A_j$ is the hyperfine coupling tensor for radical $j$.
Following the minimal model, the two radicals are assigned one anisotropic and one isotropic, \cite{Rit00707}
\begin{align}
 \bm A_1&=
 \begin{pmatrix}
 A_1&0&0\\
 0&A_1&0\\
 0&0&0
 \end{pmatrix},
 &
 \bm A_2&=A_2\mathbf I_{3\times 3}.
 \label{eq:Atensors}
\end{align}
The magnetic field is described by
\begin{align}
 \bm B=B(\sin\theta\cos\phi,\sin\theta\sin\phi,\cos\theta).
 \label{eq:Bangles}
\end{align}
As $\bm A_1$ is axially symmetric, $\phi$ can be fixed to
zero without loss of generality.

The radical pair is initialized in an electronic singlet state with unpolarized
nuclear spins,
\begin{align}
 \hat\rho(0)
 =\ket{S}\bra{S}\otimes\frac{{\bf I}_{\rm nuc}}{N_{\rm nuc}},
 \qquad N_{\rm nuc}=4,
 \label{eq:rho0}
\end{align}
where
\begin{align}
 \ket{S}=\frac{1}{\sqrt2}
 (\ket{\uparrow\downarrow}-\ket{\downarrow\uparrow}).
\end{align}
The projection operators, extended over the nuclear space, are
\bsube
\begin{align}
\hat P_S &= \ket{S}\bra{S}\otimes {\bf I}_{\rm nuc},
\\
\hat P_T &= \left(
\ket{T_+}\bra{T_+}
+\ket{T_0}\bra{T_0}
+\ket{T_-}\bra{T_-}
\right)\otimes {\bf I}_{\rm nuc},
\end{align}
\esube
where $\ket{S}$ and $\ket{T_+}$, $\ket{T_0}$, $\ket{T_-}$ are respectively the electronic singlet and triplet states.
The reaction superoperator for recombination is\cite{Rit00707}
\begin{align}
 \mathcal K_{\rm rec}\hat O
 =\frac{k_S}{2}\anticomm{\hat P_S}{\hat O}
 +\frac{k_T}{2}\anticomm{\hat P_T}{\hat O}.
 \label{eq:Krec}
\end{align}
Within the closed-system approximation,
\begin{align}
 \frac{{\rm d}\hat\rho}{\ t}
 =-\frac{i}{\hbar}[\hat H_{\rm RP},\hat\rho]
 -\mathcal K_{\rm rec}\hat\rho.
 \label{eq:Haberkorn}
\end{align}
The singlet and triplet populations are
\begin{align}
 P_\alpha(t)={\rm Tr}[\hat P_\alpha\hat\rho(t)],
 \qquad \alpha\in\{S,T\}.
 \label{eq:populations}
\end{align}
The reaction yields are defined as
\begin{align}\label{Pht}
    \Phi_\alpha(B,\theta)=
   \int_0^\infty {\rm d}\tau\,  k_\alpha P_\alpha(\tau).
\end{align}


\subsection{System--bath Hamiltonian and bath correlation functions}
\label{sec:sys_bath}

The radical pair and the explicitly retained nuclear spins constitute the
reduced system.  The unresolved environmental degrees of freedom are divided
into a spin bath, denoted by ``${\rm sp}$'', and a bosonic
bath, denoted by ``${\rm bo}$''.  We assume that the two baths are initially in
thermal equilibrium and are mutually independent.  The total Hamiltonian is
written as
\begin{align}
    \hat H_{\rm T}
    =&\,\hat H_{\rm RP}
    +\hat H_{\rm RP-sp}+\hat H_{\rm RP-bo} +\hat h_{\rm sp}+\hat h_{\rm bo},
    \label{eq:HT_mixed_group}
\end{align}
where
\bsube
\begin{align}
    \hat H_{\rm RP-sp}
    =&\,\sum_{a=1}^2
    \hat Q^{\rm sp}_a \hat F^{\rm sp}_a, \qquad \hat Q^{\rm sp}_a=\hat S_{az}, 
    \label{eq:H_Ssp_group}\\
    \hat H_{\rm RP-bo}
    =&\,\hat Q^{\rm bo}\hat X^{\rm bo}, \qquad
    \hat Q^{\rm bo}=\hat P_S-\hat P_T .
    \label{eq:H_Sbo_group}
\end{align}
\esube
Here the system operators are dimensionless, whereas
$\hat F_a^{\rm sp}$ and $\hat X_a^{\rm bo}$ have energy units.
The influence of bosonic bath is taken to account for the energy 
fluctuation between
singlet and triplet states.
Its  influence is entirely described by the bath correlation functions
\begin{align}
C^{\rm bo}(t)\equiv\left\langle \hat X^{\rm bo}(t)\hat X^{\rm bo}(0) \right\rangle_{\rm bo}
\end{align}
with 
\begin{align}
\hat X^{\rm bo}(t)\equiv e^{i\hat h_{\rm bo}t/\hbar}\hat X^{\rm bo}e^{-i\hat h_{\rm bo}t/\hbar}.
\end{align}
 Here, $\langle \cdot \rangle_{\rm bo} = {\rm Tr}(\cdot\, \hat\rho_{\rm bo}^{\rm eq})$. The thermal
boson-bath state is
$\hat\rho_{\rm bo}^{\rm eq}=e^{-\beta\hat h_{\rm bo}}/{\rm Tr}_{\rm bo}e^{-\beta\hat h_{\rm bo}}$ with $\beta=(k_{\rm B}T)^{-1}$.
The correlation function is related to the bononic bath spectral density $J^{\rm bo}(\omega)$ 
via the fluctuation--dissipation theorem as
\begin{equation}\label{fdt}
C^{\rm bo}(t)=\frac{\hbar}{\pi}\int_{-\infty}^{\infty}\!{\rm d}\omega\,
  \frac{e^{-i\omega t}J^{\rm bo}(\omega)}{1-e^{-\beta \hbar \omega}}.
\end{equation}

We futher consider the spin bath represented by independent spin-$1/2$ modes,
\cite{Yin24144112}
$\hat h_{\rm sp}=\hbar\sum_{\ell}
    \omega_\ell \hat s_\ell^z$ and 
$\hat F^{\rm sp}_a=\sqrt{2}\sum_{\ell}    c_{a\ell}\hat s_\ell^x$.
The bath spin operators are dimensionless and satisfy
    $[\hat s^\alpha_\ell,\hat s^\beta_{\ell'}]
    =i\delta_{\ell\ell'}\epsilon_{\alpha\beta\gamma}
    \hat s^\gamma_\ell.$
The spin bath correlation function is defined as
\begin{equation}
    C^{\rm sp}_{ab}(t)
    \equiv \left\langle
    \hat F^{\rm sp}_a(t)\hat F^{\rm sp}_b(0)
    \right\rangle_{\rm sp},
    \label{Cmunu_def_group}
\end{equation}
where $\hat F^{\rm sp}_a(t)\equiv 
e^{i\hat h_{\rm sp}t/\hbar}\hat F^{\rm sp}_a
e^{-i\hat h_{\rm sp}t/\hbar}$, $\langle \cdot \rangle_{\rm sp} = {\rm Tr}(\cdot\, \hat\rho_{\rm sp}^{\rm eq})$. The thermal
spin-bath state is
$\hat\rho_{\rm sp}^{\rm eq}=e^{-\beta\hat h_{\rm sp}}/{\rm Tr}_{\rm sp}e^{-\beta\hat h_{\rm sp}}$.
A spin bath is generally non-Gaussian. However, when $N_{\rm sp}\to\infty$
with central-limit scaling $c_{a\ell}\sim N_{\rm sp}^{-1/2}$, the higher order
cumulants vanish.
\cite{Sua919115,Mak992823}

The correlation function in \Eq{Cmunu_def_group} can be obtained via the fluctuation--dissipation theory (FDT) for the spin bath,\cite{Yin24144112}
\begin{align} \label{fdt_sp}
C^{\rm sp}_{ab}(t) = \frac{\hbar}{\pi} \int_{-\infty}^{\infty} {\rm d} \omega \, \frac{e^{-i\omega t} J^{\rm sp}_{ab}(\omega)}{1+e^{-\beta \hbar \omega}}.
\end{align}
The bare spin-bath spectral-density matrix is
\begin{align}
 J_{ab}^{\rm sp}(\omega)
&= \frac{1}{2\hbar} \int_{-\infty}^{\infty} {\rm d}t\,
e^{i\omega t}
\left\langle
\left\{ \hat F^{\rm sp}_{a}(t), \hat F^{\rm sp}_{b}(0) \right\}
\right\rangle_{\rm sp}
\nonumber\\
&= \frac{\pi}{2\hbar}\sum_\ell c_{a\ell}c_{b\ell}
    [\delta(\omega-\omega_\ell) + \delta(\omega+\omega_\ell)].
\end{align}
Equation (\ref{fdt_sp}) is equivalent to
\begin{align}\label{fdt_bo}
C_{ab}^{\rm sp}(t) = \frac{\hbar}{\pi} \int_{-\infty}^{\infty} {\rm d} \omega \, \frac{e^{-i\omega t} J^{\rm sp;eff}_{ab}(\omega;\beta)}{1-e^{-\beta\hbar\omega}}
 \end{align}
with
\begin{equation}
    J^{\rm sp;eff}_{ab}(\omega;\beta)
    =J_{ab}^{\rm sp}(\omega)
    \tanh\left(\frac{\beta\hbar\omega}{2}\right) .
    \label{Jeff_group}
\end{equation}
It should be noted again that the spin bath is not Gaussian in general. 
However, in the thermodynamic limit when $N_{\rm sp}\to\infty$ and the coupling coefficients scaled as 
\(c_{a\ell}\sim N_{\rm sp}^{-1/2}\), all cumulants higher than second order vanish, 
leading to the linear response condition.\cite{Sua919115,Mak992823}
The bath influence  therefore behaves Gaussian and is completely described by the  correlation functions.
Under this condition, the spin bath can be further represented by an effective Gaussian bosonic environment [\Eq{fdt_bo}] with the temperature-dependent effective spectral density \(J_{ab}^{\rm sp;eff}(\omega;\beta)\) [\Eq{Jeff_group}].\cite{Yin24144112} 
This mapping is an equivalence at the level of the reduced system dynamics influenced via bath correlation functions. In later numerical simulations, we will apply the non-Markovian and non-perturbative HEOM/DEOM propagation with formalism presented in \Sec{mixedbathdeom}
to explore the bath memory effect.

\section{Open-system descriptions with simultaneous spin and bosonic environments}
\label{thsec3}

\subsection{HEOM/DEOM for simultaneous non-Markovian spin and bosonic baths}
\label{mixedbathdeom}

The common construction of HEOM/DEOM is on the basis of exponential series expansion of bath correlation functions.
The effective-Gaussian spin-bath correlation and the bosonic
correlation function are decomposed independently as 
\bsube
\begin{align}
    C^{\rm sp}_{ab}(t)
    &\simeq\sum_{\kappa=1}^{K^{\rm sp}}
    \eta^{\rm sp}_{ab\kappa}
    e^{-\gamma^{\rm sp}_{\kappa}t},
    \label{eq:Csp_expansion_group}\\
    C^{\rm bo}(t)
    &\simeq\sum_{k=1}^{K^{\rm bo}}
    \eta^{\rm bo}_{k}e^{-\gamma^{\rm bo}_{k}t}.
    \label{eq:Cbo_expansion_group}
\end{align}
\esube
In practice, such exponential series expansion can be obtained via the contour integral of the Fourier integrand in FDT [\Eq{fdt} or \Eq{fdt_bo}] by a sum-over-poles decomposition\cite{Din12224103} or by a time-domain Prony fitting scheme.\cite{che22221102}
For the present radical-pair problem, we assume all exponents are real and independent of coupling mode label. Occasions of complex conjugate pairs can refer to Refs.\,\onlinecite{Yan14054105,Xu151816,Wan22170901}.

The HEOM/DEOM formalism for the radical-pair problem with simultaneous non-Markovian spin and bosonic bath is constructed as follows.\cite{Tan906676,Yan04216,Xu05041103,Shi09084105,Shi09164518,Tan20020901,Yan14054105,Xu151816,Wan22170901,Yin24144112}
\begin{align}
    \!\!\!\dot{\hat\rho}_{\mathbf n,\mathbf m}^{(N)}
    =&-\Big(
    \frac{i}{\hbar}{\cal L}_{\rm RP}+{\cal K}_{\rm rec}
    +\sum_{a\kappa}n_{a\kappa}
    \gamma^{\rm sp}_{\kappa}
    +\sum_{k}m_{k}\gamma^{\rm bo}_{k}
    \Big)\hat\rho_{\mathbf n,\mathbf m}^{(N)}\notag\\
    &-\frac{i}{\hbar}\sum_{a\kappa}
    {\cal A}^{\rm sp}_{a \kappa}
    \hat\rho_{\mathbf n^{+}_{a\kappa},\mathbf m}^{(N+1)}
    -\frac{i}{\hbar}\sum_{a\kappa}
    {\cal C}^{\rm sp}_{a\kappa}
    \hat\rho_{\mathbf n^{-}_{a\kappa},\mathbf m}^{(N-1)}\notag\\
    &-\frac{i}{\hbar}\sum_{k}{\cal A}^{\rm bo}_{k}
    \hat\rho_{\mathbf n,\mathbf m^{+}_{k}}^{(N+1)}
    -\frac{i}{\hbar}\sum_{k}
    {\cal C}^{\rm bo}_{k}
    \hat\rho_{\mathbf n,\mathbf m^{-}_{k}}^{(N-1)} ,
    \label{eq:heom_deom_mixed_group}
\end{align}
where \({\cal L}_{\rm RP}\hat O\equiv[\hat H_{\rm RP},\hat O],\)
the tier-up superoperators
\bsube
\begin{align}
    & {\cal A}^{\rm sp}_{a\kappa} \hat O \equiv \sqrt{(n_{a\kappa}+1)|\eta_{aa\kappa}^{\rm sp}|} [\hat Q^{\rm sp}_a,\hat O],  
     \\
    & {\cal A}^{\rm bo}_{k}\hat O \equiv \sqrt{(m_{k}+1)|\eta_{k}^{\rm bo}|} [\hat Q^{\rm bo},\hat O],
    \label{eq:superA_mixed_group}
\end{align}
\esube
and the tier-down superoperators
\bsube
\begin{align}
    {\cal C}^{\rm sp}_{a\kappa}\hat O
    \equiv& \sqrt{n_{a \kappa}/|\eta_{aa\kappa}^{\rm sp}|} \sum_{b}\, \Big(\eta^{\rm sp}_{ab\kappa}
    \hat Q^{\rm sp}_b\hat O
    -\eta^{\rm sp\,\ast}_{ab\kappa}
    \hat O\hat Q^{\rm sp}_b\Big),
    \label{eq:superC_sp_group}\\
    {\cal C}^{\rm bo}_{k}\hat O
    \equiv&\,\sqrt{m_{k}/|\eta_{k}^{\rm bo}|} (\eta^{\rm bo}_{k}\hat Q^{\rm bo}\hat O
    -\eta^{\rm bo\,\ast}_{k}\hat O\hat Q^{\rm bo}) .
    \label{eq:superC_bo_group}
\end{align}
\esube
The multiple indices are defined by
\begin{align*}
{\bf n}&\equiv\{n_{a \kappa}\geq 0;\,\kappa=1,\cdots,K^{\rm sp}; a =1,2 \},
\\
{\bf m}&\equiv\{m_k\geq 0;\,k=1,\cdots,K^{\rm bo}\},
\end{align*}
and $\mathbf n^{\pm}_{a\kappa}$ and
$\mathbf m^{\pm}_{k}$ differ from $\mathbf n$ and $\mathbf m$ only by $n_{a\kappa} \pm 1$ and
$m_{k} \pm 1$. 
The $\hat \rho_{\rm S}=\hat\rho_{\mathbf 0,\mathbf 0}^{(0)}$ is the reduced density operator of the radical-pair system.
For the
factorized initial state, \(\hat\rho_{\rm T}(0)=\hat\rho_{\rm S}(0)
    \hat\rho_{\rm sp}^{\rm eq}\hat\rho_{\rm bo}^{\rm eq}\),
all higher-tier operators vanish initially.
The hierarchy of \Eq{eq:heom_deom_mixed_group} may be truncated by a total-tier condition,
\(N\equiv
\sum_{a\kappa}n_{a\kappa}+\sum_{k}m_{k}\le L\),
or by separate upper bound $L_{\rm sp}$ and $L_{\rm bo}$. 
We can also apply the hierarchical on-the-fly filtering propagator for numerical efficiency.\cite{Shi09084105,Shi09164518}
Convergence of reaction
yields and angular contrasts to be demonstrated will be checked with respect to both correlation
expansions and hierarchy truncations.

\subsection{Lindblad master equation for the Markovian reference}

We start from the time-local quantum dissipation equation,
constructed via the cumulant partial ordering 
prescription,\cite{Xu029196}
\begin{align}
 \dot {\hat\rho}_{\rm S}= &-\Big(\frac{i}{\hbar}{\cal L}_{\rm RP}+{\cal K}_{\rm rec}\Big) {\hat\rho}_{\rm S}
 \nonumber \\ 
 &-\frac{1}{\hbar}
 \sum_{a}\big[\hat Q^{\rm sp}_a,
  \widetilde Q^{\rm sp}_a\hat \rho_{\rm S}-\hat\rho_{\rm S}(\widetilde Q_{a}^{\rm sp})^{\dg}\big]
 \nonumber \\ 
 & -\frac{1}{\hbar}
 \big[\hat Q^{\rm bo},
  \widetilde Q^{\rm bo}\hat \rho_{\rm S}-\hat\rho_{\rm S}(\widetilde Q^{\rm bo})^{\dg}\big]
\label{csqdt}
\end{align}
with
\bsube
\begin{align}
\widetilde Q^{\rm sp}_a&\equiv\sum_{b} \widetilde C^{\rm sp}_{ab}(-{\cal L}_{\rm RP})\hat Q^{\rm sp}_{b},  \\
\widetilde Q^{\rm bo}&\equiv \widetilde C^{\rm bo}(-{\cal L}_{\rm RP})\hat Q^{\rm bo},
\end{align}
\esube
and
\bsube
\begin{align}
\widetilde C^{\rm sp}_{ab}(\omega)&\equiv\frac{1}{\hbar}
\int^{\infty}_{0}{\rm d}\tau\, e^{i\omega \tau}C^{\rm sp}_{ab}(\tau)\equiv C^{\rm sp}_{ab}(\omega)+iD^{\rm sp}_{ab}(\omega),  \\
\widetilde C^{\rm bo}(\omega)&\equiv\frac{1}{\hbar}
\int^{\infty}_{0}{\rm d}\tau\, e^{i\omega \tau}C^{\rm bo}(\tau)\equiv  C^{\rm bo}(\omega)+i D^{\rm bo}(\omega).
\end{align}
\esube
Note that 
$C^{\rm sp}_{ab}(\omega)=C^{\rm sp\,\ast}_{ba}(\omega)$ and $D^{\rm sp}_{ab}(\omega)=D^{\rm sp\,\ast}_{ba}(\omega)$
are the Hermite and anti-Hermite components rather than real and imaginary parts. In the following, we assume they are real.

Let
\(\hat H_{\rm RP}\ket{m}=E_m\ket{m}\) and \(
 \omega_{mn}=(E_m-E_n)/{\hbar}\) 
to decompose
\begin{align}
&\hat Q^{\rm sp}_a=
\sum_{mn}Q^{\rm sp}_{amn}|m\ra\la n|,
\quad \hat Q^{\rm bo}=
\sum_{mn}Q^{\rm bo}_{mn}|m\ra\la n|,
\nonumber\\  \nonumber
&\widetilde Q^{\rm sp}_{a}=
\sum_{bmn}\widetilde C^{\rm sp}_{ab}(\omega_{nm})Q^{\rm sp}_{bmn}|m\ra\la n|,
\nonumber\\  \nonumber
&\widetilde Q^{\rm bo}=
\widetilde C^{\rm bo}(\omega_{nm})Q^{\rm bo}_{mn}|m\ra\la n|.\nonumber
\end{align}
Define $\hat S^{\rm sp}_{amn}\equiv Q^{\rm sp}_{amn}|m\ra\la n|$
and $\hat S^{\rm bo}_{mn}\equiv Q^{\rm bo}_{mn}|m\ra\la n|$
satisfying
\(
 \hat S^{\rm sp\,\dg}_{amn}=\hat S^{\rm sp}_{anm}
\)
and
\(
 \hat S^{\rm bo\,\dg}_{mn}=\hat S^{\rm bo}_{nm}.
\)
Substituting the above in \Eq{csqdt}
and applying the rotating wave approximation,
we obtain
\begin{align}
 \dot {\hat\rho}_{\rm S}= 
 &-
 \Big(\frac{i}{\hbar}{\cal L}_{\rm RP}+{\cal K}_{\rm rec}\Big) {\hat\rho}_{\rm S}
\nonumber \\ 
&+
\frac{1}{\hbar}\sum_{abmn}
\Big[2C^{\rm sp}_{ab}(\omega_{nm})
 \hat S^{\rm sp}_{amn}{\hat\rho}_{\rm S}\hat S^{\rm sp\,\dg}_{bmn}
\nonumber
\\
&\qquad\qquad-\widetilde C^{\rm sp}_{ab}(\omega_{nm})
\hat S^{\rm sp\,\dg}_{amn}
\hat S^{\rm sp}_{bmn}
{\hat\rho}_{\rm S}
\nonumber
\\
&\qquad\qquad-\widetilde C^{\rm sp\,\ast}_{ab}(\omega_{nm})
{\hat\rho}_{\rm S}
\hat S^{\rm sp\,\dg}_{amn}
\hat S^{\rm sp}_{bmn}
\Big]
\nonumber \\ 
&+
\frac{1}{\hbar}\sum_{mn}
\Big[2C^{\rm bo}(\omega_{nm})
 \hat S^{\rm bo}_{mn}{\hat\rho}_{\rm S}\hat S^{\rm bo\,\dg}_{mn}
\nonumber
\\
&\qquad\qquad-\widetilde C^{\rm bo}(\omega_{nm})
\hat S^{\rm bo\,\dg}_{mn}
\hat S^{\rm bo}_{mn}
{\hat\rho}_{\rm S}
\nonumber
\\
&\qquad\qquad-\widetilde C^{\rm bo\,\ast}(\omega_{nm})
{\hat\rho}_{\rm S}
\hat S^{\rm bo\,\dg}_{mn}
\hat S^{\rm bo}_{mn}
\Big].
\label{csqdt2}
\end{align}
Introduce the spin-bath induced 
Lamb-shift Hamiltonian $\hat H_{\rm LS}$ and the 
bosonic-bath induced renormalization Hamiltonian $\hat H_{\rm RE}$, respectively, as
\be
\hat H_{\rm LS}=
\sum_{abmn}
D^{\rm sp}_{ab}(\omega_{nm})
\hat S^{\rm sp\,\dg}_{amn}
\hat S^{\rm sp}_{bmn}\,,
\ee
\be
\hat H_{\rm RE}=
\sum_{mn}
D^{\rm bo}(\omega_{nm})
\hat S^{\rm bo,\dg}_{mn}
\hat S^{\rm bo}_{mn}\,.
\ee
Equation (\ref{csqdt2}) can now be recast as
\begin{align}
 \dot {\hat\rho}_{\rm S}= 
 &-
 \frac{i}{\hbar}[\hat H_{\rm RP}+\hat H_{\rm LS}+\hat H_{\rm RE},{\hat\rho}_{\rm S}]-{\cal K}_{\rm rec} {\hat\rho}_{\rm S}
\nonumber \\ 
&+
\frac{1}{\hbar}\sum_{abmn}
C^{\rm sp}_{ab}(\omega_{nm})\Big[2
 \hat S^{\rm sp}_{amn}{\hat\rho}_{\rm S}\hat S^{\rm sp\,\dg}_{bmn}
\nonumber
\\
&\qquad-
\hat S^{\rm sp\,\dg}_{amn}
\hat S^{\rm sp}_{bmn}
{\hat\rho}_{\rm S}
-
{\hat\rho}_{\rm S}
\hat S^{\rm sp\,\dg}_{amn}
\hat S^{\rm sp}_{bmn}
\Big]
\nonumber \\ 
&+
\frac{1}{\hbar}\sum_{mn}
C^{\rm bo}(\omega_{nm})\Big[2
 \hat S^{\rm bo}_{mn}{\hat\rho}_{\rm S}\hat S^{\rm bo\,\dg}_{mn}
\nonumber
\\
&\qquad-
\hat S^{\rm bo\,\dg}_{mn}
\hat S^{\rm bo}_{mn}
{\hat\rho}_{\rm S}
-{\hat\rho}_{\rm S}
\hat S^{\rm bo\,\dg}_{mn}
\hat S^{\rm bo}_{mn}
\Big].
\label{csqdt3}
\end{align}
This is rightly the standard form of Lindblad master equation.\cite{Lin76119,Gor76821}
It will be used in the next section as a reference without bath memory effect,
to compare with the exact HEOM/DEOM approach.

\begin{table}[t]
\caption{Radical-pair model parameters.}
\label{tab:parameters}
\centering
\small
\begin{tabular}{@{}ccc@{}}
\toprule
\parbox[t]{0.17\textwidth}{\textbf{Parameter}} &
\parbox[t]{0.22\textwidth}{\textbf{Value}} \\
\midrule

\parbox[t]{0.17\textwidth}{$B$ [\Eq{eq:Bangles}]} &
\parbox[t]{0.22\textwidth}{$0.5\,\mathrm{G}$}
\\
\parbox[t]{0.17\textwidth}{$\gamma_e$ [\Eq{eq:HRP}]} &
\parbox[t]{0.22\textwidth}
{$17.6\,\mathrm{\mu s^{-1}}\mathrm{G^{-1}}$}
\\
\parbox[t]{0.17\textwidth}{$A_1$ [\Eq{eq:Atensors}]}  &
\parbox[t]{0.22\textwidth}{$10\,\mathrm{G}$} &
\\
\parbox[t]{0.17\textwidth}{$ A_2$ [\Eq{eq:Atensors}]}  &
\parbox[t]{0.22\textwidth}{$5\,\mathrm{G}$} &
\\
\parbox[t]{0.17\textwidth}{$k_S,k_T$ [\Eq{eq:Krec}]} &
\parbox[t]{0.22\textwidth}{$k_S=k_T=1\,\mu\mathrm{s}^{-1}$} &
\\
\bottomrule
\end{tabular}
\end{table}

\section{Numerical demonstrations}
\label{thsec4}

For the radical-pair model, the parameters used in the numerical simulations are listed in Table~\ref{tab:parameters}. These parameters are chosen following Ref.~\onlinecite{Rit00707}. The  temperature is set to $T=300 K$, corresponding to $k_{\mathrm{B}}T/\hbar=3.93\times10^{7}~\mu\mathrm{s}^{-1}$.
Obviously, it is an extremely high-temperature limit case,  that is $\beta\rightarrow 0$ in Eqs.~\eqref{fdt} and \eqref{fdt_sp}.
As a result, noting that $J^{\rm bo}(\omega)$ and
$J_{ab}^{\rm sp}(\omega)$ are odd and even functions of $\omega$,
respectively, in Eqs.~\eqref{fdt} and \eqref{fdt_sp}, the imaginary
contributions vanish, leaving only the real parts.

The boson-bath and spin-bath spectral densities are, respectively, chosen as
\begin{align}
    J^{\rm bo}_{}(\omega)
    &=\frac{\hbar\Delta^{\rm bo} (\omega/\omega^{\rm bo}_c)}{1+(\omega/\omega^{\rm bo}_c)^2}\,,  \label{drudebo}   
\\  \label{drudesp}
    J_{ab}^{\rm sp}(\omega)&=
    \frac{ \sigma_{ab} \hbar\Delta^{\rm sp}}{1+(\omega/\omega_c^{\rm sp})^2}\,. 
\end{align}
Here, $\Delta^{\rm sp}$ and $\Delta^{\rm bo}$ characterize coupling strengths 
between system and baths, whereas  $\omega_c^{\rm sp}$ and $\omega_c^{\rm bo}$ are the  cut-off frequencies related to the environmental memories.
For the two dissipative modes of the spin bath, the parameter
$\sigma_{ab}$ characterizes the correlation between their environmental
fluctuations:
\begin{equation}
    \sigma_{11}=\sigma_{22}=1,
    \qquad
    -1\leq \sigma_{12}=\sigma_{21}\leq 1 .
    \label{eq:correlation_eta_group}
\end{equation}
We will demonstrate three cases for comparison with \(\sigma_{12}=0\), \(\sigma_{12}=1\), and \(\sigma_{12}=-1\) corresponding respectively to independent, fully correlated, and fully anti-correlated environmental fluctuations between the two coupling modes.  
Substituting Eqs.~\eqref{drudebo} and
\eqref{drudesp} into Eqs.~\eqref{fdt} and
\eqref{fdt_sp}, respectively, and taking the high-temperature limit
$\beta\rightarrow 0$, we obtain
\begin{equation}
C^{\rm bo}(t)
\simeq
\hbar\Delta^{\rm bo}
k_{\rm B}T
e^{-\omega_c^{\rm bo}t} \equiv \eta^{\rm bo} e^{-\omega_c^{\rm bo}t},
\label{eq:realC_lorentz_group_boson}
\end{equation}
and
\begin{equation}
C_{ab}^{\rm sp}(t)
\simeq
\frac{1}{2}
\sigma_{ab}\hbar^2
\Delta^{\rm sp}\omega_c^{\rm sp}
e^{-\omega_c^{\rm sp}t} \equiv \sigma_{ab} \eta^{\rm sp}\,e^{-\omega_c^{\rm sp}t} ,
\label{eq:realC_lorentz_grou_spin}
\end{equation}
for the bosonic and spin  environments, respectively.
In the following demosntrations, we will fix $\eta^{\rm sp}/\hbar^2
=6\mu\mathrm{s}^{-2}$, 
$\eta^{\rm bo}/\hbar^2
=1\mu\mathrm{s}^{-2}$
and $\omega_c^{\rm bo}=1\mu\mathrm{s}^{-1}$ while vary $\omega_c^{\rm sp}=90,\,6,\,0.2\mu\mathrm{s}^{-1}$ to concentrate on the memory effect of the spin bath.

\begin{figure}
    \centering
    \includegraphics[width=1\linewidth]{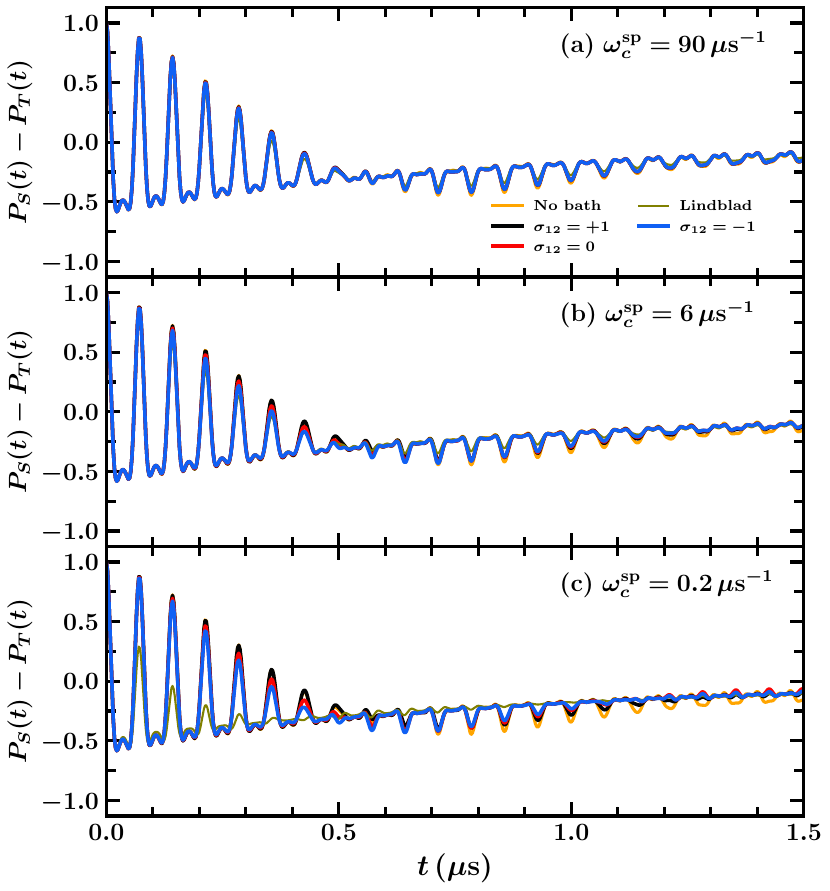}
    \caption{Time evolution of the singlet--triplet population in term of
$P_S(t)-P_T(t)$ for the spin-bath cutoff frequencies
$\omega_c^{\rm sp}=90$, $6$, and $0.2~\mu{\rm s}^{-1}$ in three panels, respectively.
In each panel the DEOM results are shown with thick lines for fully correlated
($\sigma_{12}=+1$), uncorrelated ($\sigma_{12}=0$), and fully
anticorrelated ($\sigma_{12}=-1$) spin-bath fluctuations. The no-bath
result and the Markovian Lindblad result for $\sigma_{12}=0$ are
included (shown in thin lines) for comparison. The angle is fixed at $\theta=68^{\circ}$. The remaining parameters are given in
Table~\ref{tab:parameters} and main text.
    }
    \label{fig1}
\end{figure}

\begin{figure}
    \centering
    \includegraphics[width=1\linewidth]{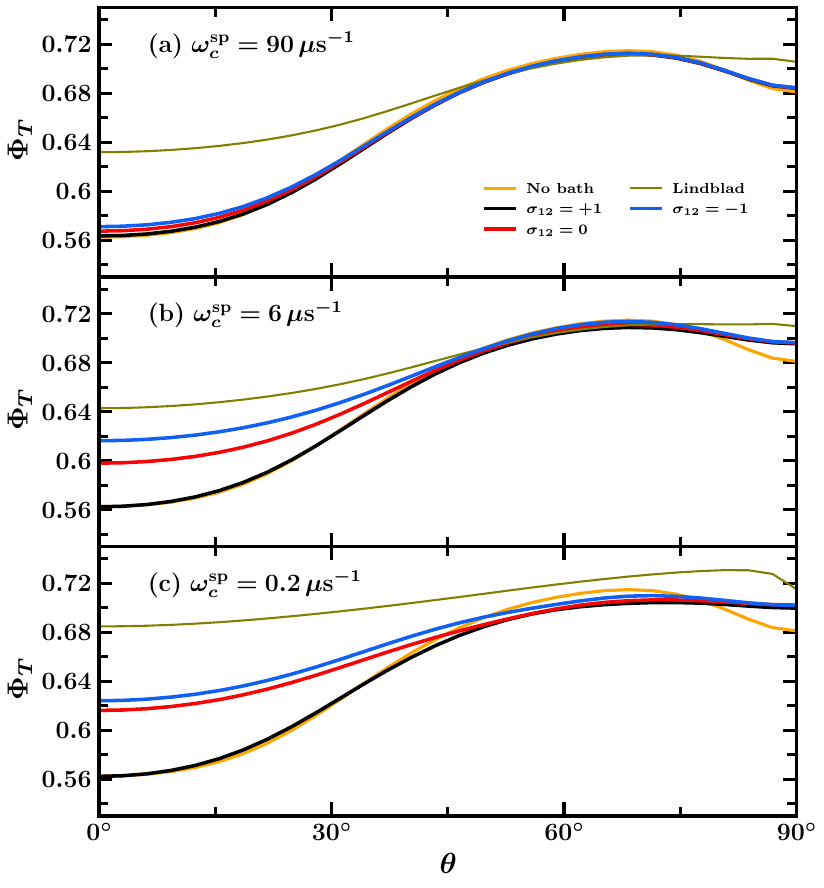}
    \caption{Angular dependence of the triplet reaction yield $\Phi_T$ in \Eq{Pht}
for the spin-bath cutoff frequencies $\omega_c^{\rm sp}=90$, $6$, and
$0.2~\mu{\rm s}^{-1}$in three panels, respectively.
In each panel the DEOM results are shown with thick lines for fully correlated
($\sigma_{12}=+1$), uncorrelated ($\sigma_{12}=0$), and fully
anticorrelated ($\sigma_{12}=-1$) spin-bath fluctuations. The no-bath
result and the Markovian Lindblad result for $\sigma_{12}=0$ are
included (shown in thin lines) for comparison. The remaining parameters are given in
Table~\ref{tab:parameters} and main text.
    }
    \label{fig2}
\end{figure}

Figure~\ref{fig1} and Figure~\ref{fig2}  display the effects of bath memory and cross-correlation on the radical-pair dynamics and its magnetic-field sensitivity through its angular dependence. 
The triplet reaction yield $\Phi_T$ shown in Fig.~\ref{fig2} is obtained by numerical integration  from $0$ to $10~\mu{\rm s}$ according to Eq.~(\ref{Pht}).
As manifested in Fig.~\ref{fig1}, with increasing bath memory from panels (a) to (c), corresponding to a decrease in $\omega_c^{\rm sp}$, the DEOM results exhibit larger deviations from both the Lindblad and no-bath results. 
Meanwhile, in the regime where the memory effect is most pronounced [panel (c)], the influence of bath cross-correlation also becomes the most evident, leading to largest differences among  $\sigma_{12}=+1$, $0$, and $-1$ cases.

The behavior changes with $\sigma_{12}$ can be analyzed as follows. The fully correlated ($\sigma_{12}=+1$) and fully anticorrelated ($\sigma_{12}=-1$) actually correspond to the system dissipative modes $\hat S_{1z}+\hat S_{2z}$ and $\hat S_{1z}-\hat S_{2z}$, respectively.
For $\sigma_{12}=+1$, $\hat S_{1z}+\hat S_{2z}$ does not directly couple the singlet state $\ket{S}$ with the $\ket{T_0}$ state. Therefore, the bath effect on the singlet--triplet decoherence is relatively weak. 
In contrast, for $\sigma_{12}=-1$, $(\hat S_{1z}-\hat S_{2z})\ket{S}=\ket{T_0}$,  thus it can directly affect the singlet--triplet conversion.

These dynamical differences become more apparent in the integrated triplet reaction yield shown in Fig.~\ref{fig2}. All curves retain the basic angular anisotropy produced by the anisotropic hyperfine interaction: $\Phi_T$ increases from $\theta=0^\circ$, reaches a broad maximum around $65^\circ$--$75^\circ$, and then decreases slightly toward $\theta=90^\circ$. 
The only exception is the Lindblad result in panel (c), which shows a maximum near $\theta=90^\circ$. This indicates the necessity of properly treating the bath memory effect.
In Fig.~\ref{fig2}, the comparison between the HEOM/DEOM and the Lindblad results generally demonstrates that the reaction yield is particularly sensitive to non-Markovian effects. 
Even in the fast-modulation regime [panel (a)], where the transient population dynamics seems quite similar, their small differences can accumulate in the time integration defining $\Phi_T$, leading to more apparent deviations in the reaction yield.
The dependence on $\sigma_{12}$ can be understood in a similar way to that discussed for Fig.~\ref{fig1}.
Taken together, Figs.~\ref{fig1} and \ref{fig2} show that both bath memory and spatial correlation play important roles in radical-pair spin dynamics, and more crucially in  angular dependence of the reaction yield.
Therefore, a non-Markovian treatment of the correlated spin environment is required for a proper description of the compass response to the external magnetic field.

\section{Summary}
\label{thsum}

In summary, we have conducted a HEOM/DEOM simulation of a
radical-pair system simultaneously influenced by a spin environment
and a bosonic environment. 
The spin bath acts locally on the two
electron spins, whereas the bosonic bath describes fluctuations of the
singlet--triplet energy difference. In the thermodynamic and
linear-response limit, the spin bath can be represented by an
effective Gaussian environment whose influence is completely
characterized by its correlation functions. This mapping
allows the spin and bosonic environments to be treated uniformly within a nonperturbative and non-Markovian hierarchical
framework. A time-local Lindblad master equation has also been presented
as a Markovian reference. 
The numerical results demonstrate the key role of the environmental memory effect in both the dynamics and the angular dependence of the reaction yield.
These findings identify correlated
environmental memory as a crucial ingredient for radical-pair
magnetoreception. They also show that a Markovian relaxation model may fail to capture important effects even when it approximately reproduces some features of the transient dynamics.

\begin{acknowledgments}
Support from the National Natural Science Foundation of China (Grant
Nos.~22373091 and 22573099) is gratefully acknowledged.  Simulations were
performed on the robotic AI-Scientist platform of Chinese Academy of Sciences.
\end{acknowledgments}


\end{document}